\documentclass[tightenlines,11pt,twocolumn]{revtex4-1}

	\usepackage{newtxtext}
	\usepackage{newtxmath}
	\usepackage{bm}
	\usepackage{amsmath}
	\usepackage{upgreek}
	\usepackage{setspace}
	\usepackage[margin=1.0in]{geometry}
	\usepackage{graphicx}
	\usepackage{hyperref}

\newcommand{\Rb}{Rb}
\newcommand{\K}{K}
\newcommand{\ttf}{$T/T_\mathrm{F}$}
\newcommand{\ttfn}[1]{$T/T_\mathrm{F} = #1$}
\newcommand{\bUnit}{cm$^3$s$^{-1}$K$^{-1}$}
\newcommand{\uK}{$\upmu$K}
\newcommand{\tstrut}{\rule{0pt}{2.6ex}}

\begin{document}

\title{A Fermi Degenerate Gas of Polar Molecules}
\author{Luigi De Marco, Giacomo Valtolina, Kyle Matsuda, William G. Tobias, Jacob P. Covey,$^{\dagger}$ Jun Ye$^{*}$}
\affiliation{JILA, NIST and the University of Colorado, and Department of Physics, University of Colorado, 440 UCB, Boulder, CO 80309, USA.\\
$^\dagger${\normalfont Present address:} Department of Physics, California Institute of Technology, 1200 East California Blvd., Pasadena, CA 91125\\
$^*${\normalfont To whom correspondence should be addressed. e-mail: ye@jila.colorado.edu}}

\begin{abstract}
It has long been expected that quantum degenerate gases of molecules would open access to a wide range of phenomena in molecular and quantum sciences. However, the very complexity that makes ultracold molecules so enticing has made reaching degeneracy an outstanding experimental challenge over the past decade. We now report the production of a Fermi degenerate gas of ultracold polar molecules of potassium--rubidium (KRb). Through coherent adiabatic association in a deeply degenerate mixture of a rubidium Bose-Einstein condensate and a potassium Fermi gas, we produce molecules at temperatures below 0.3 times the Fermi temperature. We explore the properties of this reactive gas and demonstrate how degeneracy suppresses chemical reactions, making a long-lived degenerate gas of polar molecules a reality.
\end{abstract}

\maketitle

Ultracold polar molecules have received attention as ideal candidates to realize a plethora of proposals in molecular and many-body physics. These include the development of chemistry in the quantum regime~\cite{Carr2009}, the emulation of strongly interacting lattice spin models~\cite{Micheli2006,Osterloh2007, Buchler2007,Gorshkov2011,Baranov2012}, the production of topological phases in optical lattices~\cite{Cooper2009,Yao2013,Syzranov2015,Peter2015}, the exploration of fundamental symmetries~\cite{Kozlov1995,Flambaum2007,Hudson2011,TheACMECollaboration2014,Cairncross2017}, and the study of quantum information science~\cite{Andre2006,Yelin2006,Ni2018}. While magnetic atoms also exhibit long-ranged dipolar interactions and can be used to carry out these proposals~\cite{Lahaye2009,Aikawa2014a}, polar molecules offer more tunable, stronger interactions and additional degrees of freedom. A low-entropy, quantum degenerate sample is a prerequisite for many of these explorations.

The intrinsic complexity of molecules relative to atoms, owing to the additional rotational and vibrational degrees of freedom, has made their cooling to ultralow temperatures one of the most significant experimental challenges in molecular physics~\cite{Bohn2017a}. While the direct laser cooling of certain diatomic molecules has progressed enormously in recent times so that magneto-optic~\cite{Hummon2013,Barry2014, Anderegg2017,Truppe2017} and pure optical~\cite{Anderegg2018} trapping have been demonstrated, phase space density in these systems remains many orders of magnitude away from degeneracy. To date, by far the coldest diatomic molecules have been made by cooling atoms to a few hundred nanokelvin ($10^{-9}$~K) and coherently associating the ultracold atoms into deeply bound molecules using a Fano-Feshbach resonance~\cite{Chin2010} followed by stimulated Raman adiabatic passage (STIRAP)~\cite{Ni2008}.

Thus far, KRb~\cite{Ni2008}, NaK~\cite{Park2015a,Seeßelberg2018}, RbCs~\cite{Takekoshi2014,Molony2014}, NaRb~\cite{Guo2016}, and LiNa~\cite{Rvachov2017} have successfully been produced in deeply bound molecular states. Typically, such molecules can be produced in numbers ranging from hundreds to tens of thousands and at temperatures ranging from 250\,--\,600~nK. Reaching degeneracy in these experiments has been impeded by two major factors: the production of an adequate mixture of atoms to make a sufficient number of molecules, and rapid molecular loss. Challenges in producing a suitable mixture can be technical or physical, such as the immiscibility of two Bose-Einstein condensates (BECs)~\cite{Reichsollner2017}. Molecules can be lost due to chemical reactions; for example, KRb undergoes the exothermic $2\mathrm{KRb}\to \mathrm{K}_2 + \mathrm{Rb}_2$ reaction~\cite{Ospelkaus2010, Ni2010}. Even molecules predicted to have endothermic reactions show large inelastic loss due to the complex nature of the scattering process, which is still being investigated~\cite{Mayle2013,Park2015a,Guo2016}. Indeed, the lowest entropy samples of ground-state molecules have been produced in a three-dimensional optical lattice, where chemical reactions cannot occur, with an entropy of just $2.2k_\mathrm{B}$ per particle~\cite{Moses2015}; however, producing quantum degenerate molecules in a bulk gas remains an outstanding experimental goal.

\begin{figure*}[t]
\centering
\includegraphics[width=5.5in]{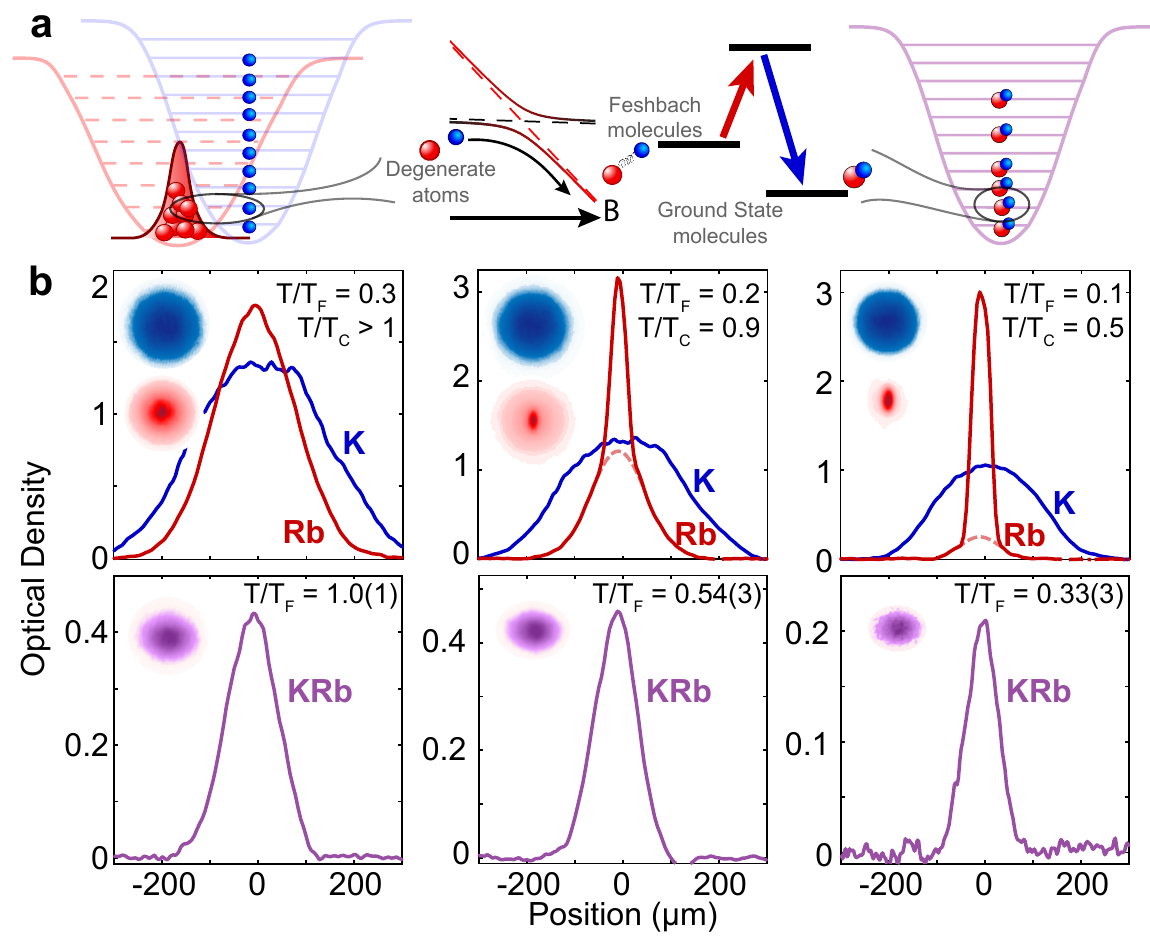}
\caption{\label{fig:figure1} \textbf{a. Production of Fermi degenerate molecules.} Beginning with degenerate gases of \Rb{} and \K{}, Feshbach molecules are created by sweeping a magnetic field through a Fano-Feshbach resonance. The weakly bound molecules are coherently transferred to the ground state using STIRAP. The resulting molecules reflect the degeneracy of their parent atoms. \textbf{b.~Degenerate gases of atoms and molecules.} Slices through images of atomic mixtures (2 averages, 16~ms TOF, upper row) and ground state molecules (4 averages, 10~ms TOF, lower row) for molecular \ttf{} ranging from 0.3 to 1. Inset images show  false-color 2D column density, with blue corresponding to \K{} and red to \Rb{}. The TOF images reflect the differing momentum distributions of the atoms and molecules.
}
\end{figure*}

\begin{table*}[t]
\centering
\begin{tabular}{c|cc|cc|cc}
$T$ (nK)& Rb Number & $T/T_\mathrm{c}$ & K Number & \ttf{} & KRb Number & \ttf{}\\ 
\hline\hline
$230$\tstrut & $6\times10^5$ & $> 1$ & $1.2\times10^6$ & 0.3 & $1.0(1)\times10^5$ & 1.0(1)\\
$110$& $2\times10^5$ & $0.9$ & $1\times10^6$ & 0.2 & $5.0(5)\times10^4$ & 0.54(3)\\
$50$& $7\times10^4$ & $0.5$ & $5\times10^5$ & 0.1 & $3.0(5)\times10^4$ & 0.33(3)
\end{tabular}
\caption{\label{tbl:tbl1} Atom and molecule conditions corresponding to Fig~\ref{fig:figure1}b.}
\end{table*}

In this paper, we report the production of $10^5$ fermionic $^{40}$K$^{87}$Rb molecules at 250~nK and as many as $2.5\times10^4$ molecules at 50~nK, the latter corresponding to \ttfn{0.3}, where $T_\mathrm{F}$ is the Fermi temperature. By generating the molecules from two deeply degenerate gases of $^{40}$K and $^{87}$Rb in a crossed optical dipole trap (xODT), we demonstrate that molecules can be produced at a range of values of \ttf{}, which depend sensitively on the initial atomic conditions. The molecular density profile is shown to be a Fermi-Dirac distribution, and we measure a corresponding deviation from the classical internal energy of the system. Finally, we demonstrate that quantum degeneracy is accompanied by a suppression of chemical reactions due to the reduction of density fluctuations in the center of the Fermi gas.

It is well understood that the efficiency of ultracold molecule production is limited at low temperatures by rapid three-body recombination of the atomic species as well as the spatial mismatch between atomic density distributions~\cite{Herbig2003, Zirbel2008, Cumby2013}. For KRb, however, in the limit where the \K{} number vastly exceeds the \Rb{} number, these effects can be mitigated, and the conversion to molecules with respect to the minority species can be high~\cite{Cumby2013}. Furthermore, if the gases are deeply degenerate, the atoms' low entropy can be inherited by the molecules, resulting in a degenerate molecular gas as illustrated in Fig.~\ref{fig:figure1}a. A large atom number before molecular association has allowed us to take this approach, and has afforded us the flexibility to produce KRb molecules over a wide range of temperatures, densities, and \ttf{}. 

After collecting $\sim$10$^9$ \Rb{} atoms and $7\times10^7$ \K{} atoms in a vapor-cell MOT, we cool the atoms to degeneracy by performing RF evaporation in an optically plugged quadrupole trap followed by evaporation in a xODT~\cite{Note1}. After optical evaporation is complete, the xODT is recompressed such that \K{} experiences harmonic trapping frequencies of $(\omega_x,\omega_y,\omega_z) = 2\uppi\times(45,250,80)$~Hz, with gravity along the $y$--direction. Trap frequencies are reduced by a factor of 0.72 and 0.79 for \Rb{} and KRb, respectively, due to differences in mass and AC polarizability. Slices through atomic column-integrated density distributions after 16~ms time of flight (TOF) for three representative conditions are shown in the upper row of Fig.~\ref{fig:figure1}b, and the corresponding numbers are given in Table~\ref{tbl:tbl1}. While a number of technical improvements have allowed us to produce a deeply degenerate mixture with a large number of atoms, a key improvement has been the implementation of $\Lambda$-enhanced gray molasses on the D$_1$ ($4^2\mathrm{S}_{1/2}\to4^2\mathrm{P}_{1/2}$) line of \K{}~\cite{RioFernandes2012} as well as the D$_2$ ($5^2\mathrm{S}_{1/2}\to5^2\mathrm{P}_{3/2}$) line of \Rb{}~\cite{Rosi2018}. 

\begin{figure}[b]
\centering

\includegraphics[width=2.6in]{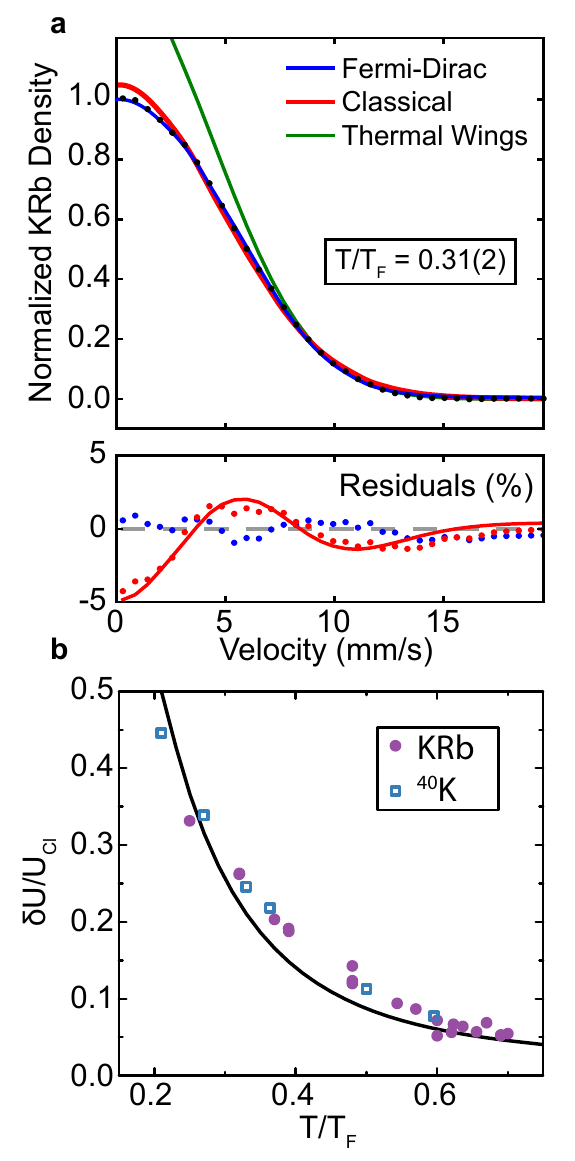}
\caption{\label{fig:figure2}\textbf{a. Fermi-Dirac distribution of molecules.} An azimuthally-averaged molecular density profile with \ttfn{0.31(2)} fit to a Fermi-Dirac distribution (blue curve) and a Gaussian (red curve). The fit residuals (lower panel) show the deviation of the Gaussian fit characteristic of Fermi degeneracy; the solid curve corresponds to the difference of the residuals. Fitting the wings of the cloud to a classical distribution (green curve) accurately captures the temperature but overestimates the density in the center. \textbf{b. Deviation from the classical energy density.} The deviation of the internal energy of the molecular gas from its classical value as \ttf{} is reduced. The solid curve is the expected result for an ideal Fermi gas, and the results for \K{} are shown for comparison.
}
\end{figure}

Ground state KRb molecules are produced by sweeping a magnetic field through an interspecies Fano-Feshbach resonance at $B=546.6$~G to produce weakly bound Feshbach molecules as illustrated in Fig.~\ref{fig:figure1}a~\cite{Ni2008}. The molecules are then coherently transferred to the ground state using STIRAP with $\sim$90\% efficiency. The difference in  trapping frequencies between molecules and atoms results in all three species having different equilibrium positions due to gravitational sag~\cite{Delannoy2001}. So while the molecules have the same initial temperature as the atoms, the initial non-equilibrium position sets the molecules in motion, and the gas rapidly heats up as potential energy is converted to kinetic energy. To mitigate this effect, prior to molecule production a one-dimensional optical lattice of 30 molecular recoil energies is turned on against gravity. This ensures that the molecules and atoms are at equilibrium at the same position in the corrugated potential. After molecules are produced, unpaired atoms are quickly blasted away with resonant light~\cite{Note1}, the lattice is then ramped off in 5~ms, and no spatial oscillations or rapid heating are observed.

By varying the initial temperature and atomic number ratio, we generate molecular gases ranging from \ttf{} greater than 1  to less than 0.3. Three representative conditions are summarized in Table~\ref{tbl:tbl1}, and slices through molecular column density distributions after 10~ms TOF are shown in the lower row of Fig.~\ref{fig:figure1}b; each molecular distribution shown is produced from the atomic conditions in the upper row of Fig.~\ref{fig:figure1}b. Over this range, the average molecular density varies from $0.5-2\times10^{12}$~cm$^{-3}$.

At low \ttf{}, the effects of degeneracy are clear on the molecular velocity distribution after TOF. Two-dimensional absorption images are collected and fit~\cite{Note1}, and Fig.~\ref{fig:figure2}a shows the azimuthally-averaged density profile of a cloud of KRb molecules after 10~ms TOF. The profile is well fit by a Fermi-Dirac distribution (blue curve, \ttfn{0.31(2)}), while the classical Gaussian distribution (red curve) overestimates the density at the center of the cloud and underestimates it in the wings. This is evident in the fit residuals, shown in the lower part of Fig.~\ref{fig:figure2}a, where the Gaussian residuals exhibit ripples that are a hallmark of Fermi degeneracy~\cite{DeMarco1999}. A Gaussian fit to the wings of the profile (green curve), where the gas looks essentially classical, captures the temperature of the molecules but deviates at the center of the cloud.

The Gaussian fit to the entire cloud systematically overestimates the temperature compared to the Fermi-Dirac fit since the Pauli exclusion principle prevents the multiple occupancy of low momentum states. The difference between the temperatures measured with the two fits is a strong indicator of the degree of quantum degeneracy~\cite{DeMarco1999}. Figure~\ref{fig:figure2}b shows the normalized difference between these, $\delta U/U_\mathrm{Cl} = 1-T/T_\mathrm{Cl}$, for KRb as a function of \ttf{}, where $T_\mathrm{Cl}$ is the temperature determined from the Gaussian fit. As \ttf{} is decreased, the normalized energy shows a deviation from the classical value, and for the most degenerate molecular clouds we currently produce, the deviation is larger than 30\%. For comparison, the same quantity is shown for \K{} at several values of \ttf{} and both show good agreement with the theoretical prediction for an ideal Fermi gas (solid line).

Given the \ttf{} measured for \K{} prior to molecular association, we expect an increase in the molecular \ttf{} by roughly a factor of 3--4 based on the change in particle number and trap frequency. However, we typically measure values of \ttf{} that are only a factor of 2.5--3 larger than that of \K{}. At values of \ttf{}~$\lesssim 0.1$ for \K{}, 85\% of the \Rb{} is condensed, and the \Rb{} to \K{} ratio is made to be roughly 1:10 (Table~\ref{tbl:tbl1}, third row) in order to minimize three-body recombination during magnetoassociation~\cite{Cumby2013,Zirbel2008a}. Under these conditions, molecules are produced with \ttf{}~$\lesssim 0.3$. Since the BEC is fairly localized to the center of the \K{} cloud (Fig.~\ref{fig:figure1}b), molecules are only produced in the lowest-entropy part of the Fermi sea~\cite{Ho2009}, resulting in molecules that have a lower \ttf{} than expected from uniform \K{} conversion over the entire distribution. At such low temperatures, the conversion from \Rb{} to Feshbach molecules can be as high as 50\%, indicating that there is good local phase space overlap between potassium and the rubidium condensate. It is the same principle of strong phase space matching that allowed for the efficient production of ground state molecules in a 3D optical lattice~\cite{Moses2015}. In contrast, at high \ttf{}, where we produce the largest number, conversion is typically 15\% of \Rb{} (Table~\ref{tbl:tbl1}, first row).

When degenerate molecules are produced, the gas is not necessarily in equilibrium due to the fact that the spatial and momentum distributions of the molecules reflect the overlapping distributions of the \Rb{} BEC and the \K{} degenerate Fermi gas. However, our observation of a Fermi-Dirac distribution (Fig.~\ref{fig:figure2}a) is consistent with a molecular gas that is close to equilibrium, and we find that degeneracy persists for the lifetime of the molecules. Furthermore, that the measured expansion energy as a function of \ttf{} (Fig.~\ref{fig:figure2}b) is consistent with that of \K{} also serves as evidence that the molecules are produced near equilibrium.   

\begin{figure}[t]
\centering

\includegraphics[width=2.6in]{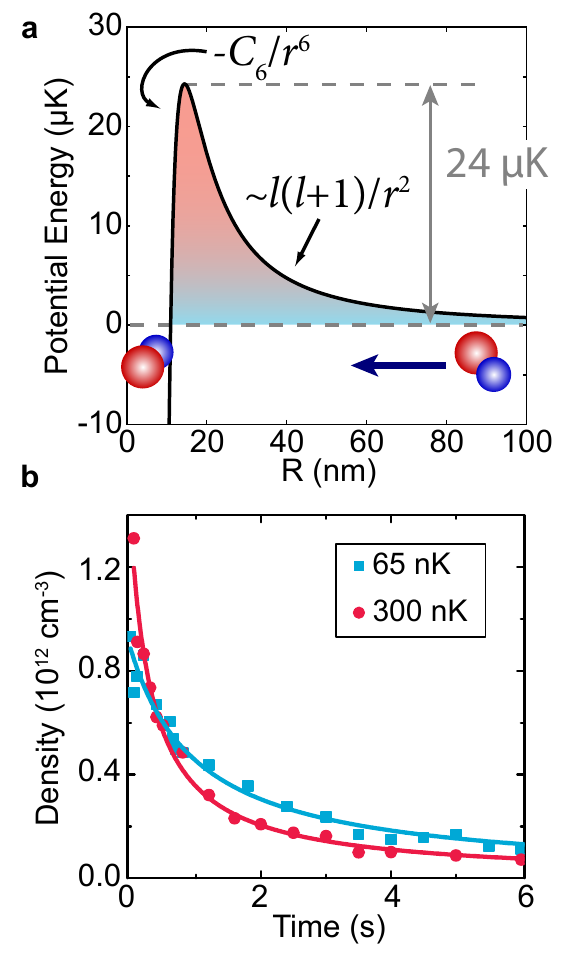}
\caption{\label{fig:figure3} \textbf{a. Intermolecular Potential.} The intermolecular $p$-wave ($l=1$) scattering potential for KRb~\cite{Ospelkaus2010}; molecules react once they have tunneled through the barrier. \textbf{b. Density Loss.} As reactions occur, molecular density decays according to a two-body rate law. The rate constant decreases with decreasing temperature in accordance with the Bethe-Wigner threshold law. 
}
\end{figure}

Given that KRb molecules are fermions, intermolecular scattering must occur in the $p$-wave channel since this is the lowest energy antisymmetric collision channel. As such, the intermolecular potential, shown in Fig.~\ref{fig:figure3}a, exhibits a centrifugal barrier through which molecules must tunnel in order to chemically react.  According to the Bethe-Wigner threshold law~\cite{Bethe1935,Wigner1948,Sadeghpour2000}, the tunneling rate (and therefore reaction rate), is proportional to the temperature, so that chemical reactions must slow down at low temperatures~\cite{Ospelkaus2010}. Examples of density loss curves and their corresponding fits (see below) for two temperatures are shown in Fig.~\ref{fig:figure3}b. At low temperature, the density decays at a slower rate compared to at high temperature.

When a molecular collision leads to a reaction, the product molecules are ejected from the trap with high energy, leaving the remaining molecules unaffected. However, collisions tend to occur in the coldest, densest part of the cloud so that the lowest energy molecules react and are lost preferentially, leading to anti-evaporation and an overall heating of the cloud. We typically observe linear heating with rates ranging from $h = $~10--30~nK/s, which are slightly larger than a simple anti-evaporation model would suggest. However, this rate is small enough so that the molecular \ttf{} remains close to its initial value over the course of the molecules' lifetime.

The reduction of density is determined by both the loss of KRb molecules as well as the increase in temperature. We fit our data to a simple two-body model that includes the effect of heating~\cite{Ospelkaus2010}, 
\begin{equation}\label{eq:densityrate}
\frac{\mathrm{d}n}{\mathrm{d}t} = -\beta n^2 - \frac{3}{2}\frac{n}{T}\frac{\mathrm{d}T}{\mathrm{d}t},
\end{equation}
where $n$ is the average classical molecular density of the bulk gas; the temperature is a measured, linear function of time, $T = T_0 + ht$; and $\beta$ is the two-body loss coefficient. Since two-particle threshold behavior predicts $\beta = bT$, the fitting of the data with Eq.~\ref{eq:densityrate} allows us to determine $b$~\cite{Note1}.

\begin{figure}[ht]
\centering
\includegraphics[width=0.9\columnwidth]{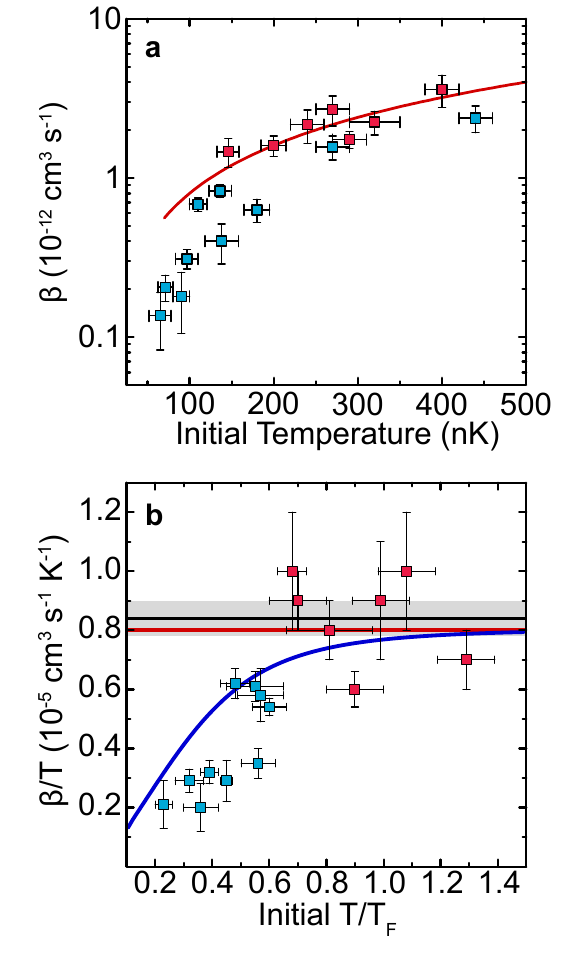}
\caption{\label{fig:figure4} \textbf{a. Reaction rate constant as a function of temperature.} The reaction rate constant $\beta$ for temperatures ranging from $T=$~70\,--\,450~nK. Blue-filled points correspond to \ttf{}~$\leq 0.6$ and red-filled points to \ttf{}~$>0.6$. The red curve is the value expected from MQDT. \textbf{b. Suppression of chemical reactions.} Temperature-normalized reaction rate constants from \textbf{a} as a function of degeneracy. The measured $\beta/T$, as determined by fitting the average density to the solution of Eq.~\ref{eq:densityrate}, appears to decrease sharply when \ttf{}~$<0.6$ due to the suppression of fluctuations. The solid black line and gray bar are the average $\beta/T$ for \ttf{}~$>0.6$ and corresponding error range. The red line is the MQDT value and the blue curve is the average relative density fluctuations.
}
\end{figure}

Measurements of $\beta$ as a function of initial temperature are shown in Fig.~\ref{fig:figure4}a. Data points with a blue face correspond to \ttf{}~$\leq0.6$ and those with a red face to \ttf{}~$>0.6$; the solid red curve is the value calculated by multi-channel quantum defect theory (MQDT)~\cite{Ospelkaus2010,Idziaszek2010}. While points with \ttf{}~$>0.6$ follow the predicted MQDT trend closely, those with \ttf{}~$\leq0.6$ show deviations at all temperatures. If instead we consider $\beta/T$, which we expect to be constant, as a function of \ttf{}, the data collapse onto a common trend independent of initial temperature. We find that at \ttf{}~$\leq0.6$, $\beta/T$ shows a strong deviation from the Bethe-Wigner threshold law, while above \ttfn{0.6}, $\beta/T$ is constant, with a measured value of $\beta/T = 0.84(6)\times10^{-5}$~\bUnit{} (black line in Fig.~\ref{fig:figure4}b,  error range shown in gray). This value is in excellent agreement with the predicted MQDT value~\cite{Ospelkaus2010,Idziaszek2010} of $\beta/T = 0.8(1)\times10^{-5}$~\bUnit{} (red line in Fig.~\ref{fig:figure4}b). Our value is somewhat lower than the previously measured value of $\beta/T = 1.2(3)\times10^{-5}$\bUnit{}~\cite{Ospelkaus2010}, with the discrepancy likely arising from the use of the corrugated potential to suppress gravitational sag in the current experiment. 

Below \ttfn{0.6}, the measured $\beta/T$ drops to values as low as $0.21(8)\times10^{-5}$~\bUnit{}. This apparent deviation from a constant value is due to the change in density correlations as the gas becomes deeply degenerate. For a classical gas, the density sets the length scale of interparticle separation, which is much larger than the deBroglie wavelength, $\Lambda$. In this case, large density fluctuations occur on the molecular scale, and two particles can easily find a configuration to scatter in the $p$-wave channel. In a Fermi degenerate gas, however, the probability of finding two molecules within a short distance of each other decreases as \ttf{} is lowered due to anti-bunching, with the average interparticle spacing being set by the deBroglie wavelength itself and ultimately by the Fermi wavevector, $2\uppi/k_\mathrm{F}$. This is the same physical phenomenon that gives rise to the Pauli pressure and the reduced compressibility of a Fermi gas. This causes an effective blockade that results in reduced density fluctuations~\cite{Rom2006,Sanner2010,Muller2010} such that $p$-wave scattering and chemical reactions are suppressed beyond the Bethe-Wigner prediction. The suppressed collision rate manifests itself within our model as a reduction in the measured $\beta/T$ for the bulk gas, though the true two-body reaction rate constant is unaffected by degeneracy for any given molecular collision. 

This effect is captured in the average relative number density fluctuation $\langle\delta n^2(\bm{r})\rangle/\langle n(\bm{r})\rangle$, which is shown as a blue line in Fig.~\ref{fig:figure4}b as a function of \ttf{}. The curve is scaled to the MQDT value of $\beta/T$ for \ttf{}~$>1$, but otherwise has no fitting parameters. That this simple consideration qualitatively describes the change in reaction rate suggests that the reduced particle fluctuations are the primary effect slowing chemical reactions. Density fluctuations are most strongly suppressed in the center of the trap~\cite{Note1} where the majority of chemical reactions occur, which is a possible explanation for why some points fall below the expected fluctuation suppression curve. Similar suppression of loss due to strong correlations has been observed in the three-body recombination rate of a one-dimensional Bose gas in which the particles have undergone fermionization~\cite{Tolra2004}. Furthermore, this effect is reminiscent of the reduction of the elastic $s$-wave collision rate observed in fermionic atoms~\cite{DeMarco2001,Aikawa2014}; however, in the elastic case, the reduction of the elastic cross section is attributed to the unavailability of empty states to scatter into, which is not relevant for inelastic collisions.

While the qualitative agreement between the data and the suppression of density fluctuations is suggestive, a complete theory must capture the subtle higher order effects that give rise to deviations from Eq.~\ref{eq:densityrate}. In particular, it is important to consider the degree to which molecules thermalize after some are lost to chemical reactions. Furthermore, as \ttf{} is decreased well below 0.3, collisions become dominated by molecules near the Fermi surface so that the mean relative collision energy deviates from the classical equipartition value to the quantum value of $\frac{3}{4}k_\mathrm{B}T_\mathrm{F}$ per particle. We expect that the true two-body loss rate will thus remain finite even at absolute zero, but the bulk Fermi gas will be increasingly stable as density fluctuations vanish at $T=0$. 

The production of a Fermi degenerate gas of dipolar molecules opens new paths in ultracold molecular science. In a bulk gas, we now have the opportunity to study chemical reactions in a regime where quantum degeneracy and quantum fluctuations compete with classical chemical reaction dynamics. Furthermore, this work shows great promise for the exploration of degenerate molecules in electric fields, where the strong dipole-dipole interaction dominates. In this limit, we expect to see interaction-induced effects such as the deformation of the Fermi surface and the development of exotic many-body correlations.

\bibliographystyle{../science}
\bibliography{../library}

\section*{Acknowledgments}
We dedicate this work to the memory of Deborah Jin, whose brilliance and passion inspire us all. We thank J. Bohn, E. Cornell, C. Greene, M. Holland, S. Moses, and A. M. Rey for useful discussions. This work is supported by NIST, AFOSR-MURI, ARO-MURI, and the NSF JILA Physics Frontier Center (NSF PHY-1734006).
\clearpage

\clearpage
\renewcommand{\thepage}{S\arabic{page}}  
\renewcommand{\theequation}{S\arabic{equation}}
\renewcommand{\thefigure}{S\arabic{figure}}
\setcounter{equation}{0}
\setcounter{page}{1}
\setcounter{figure}{0}

\section*{Supplementary Materials}

\subsection*{Materials and Methods}

\begin{figure}[ht]
\centering
\includegraphics[width=\columnwidth]{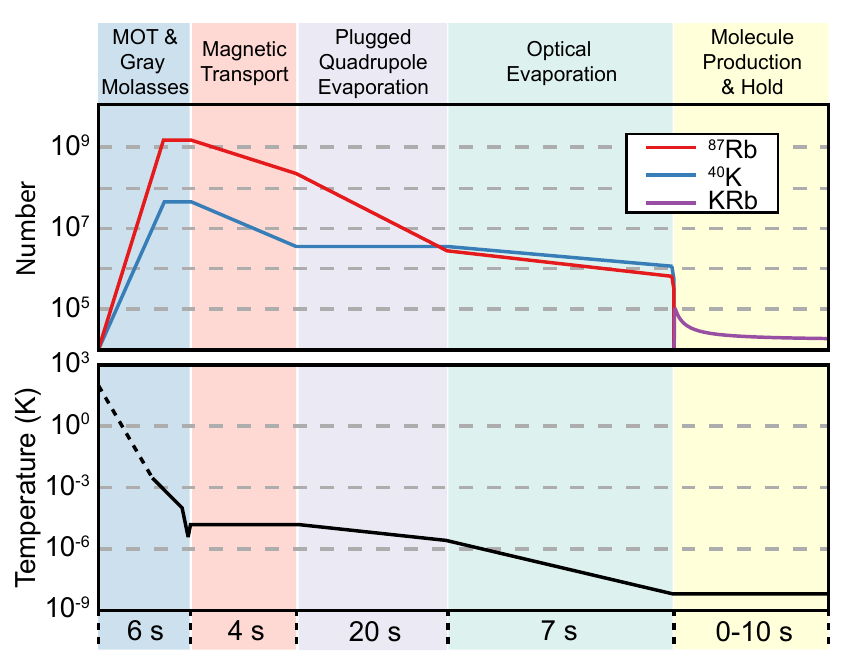}
\caption{\label{fig:figureS1} \textbf{Experimental sequence.} 
The upper panel shows the number of \Rb{}, \K{}, and KRb, while the lower panel shows the temperature of the mixture over the course of the experimental cycle. A single curve is shown for temperature since the atomic species remain thermalized. The time axis is not to scale.
} 
\end{figure}

\textbf{Preparation of degenerate atomic species.} In an atomic vapor cell ($P\sim 10^{-7}$~torr), a dual species magneto-optic trap (MOT) cools and traps \Rb{} and \K{} atoms. Once the MOT is fully loaded, a compressed MOT stage further cools the atoms, which is followed by sub-Doppler cooling. Both bright and $\Lambda$--enhanced gray molasses are sequentially performed on the $\mathrm{D}_2$ transition of \Rb{} for 2 and 8~ms, respectively, and the atoms reach a final temperature of 10~\uK{}. Simultaneously, $\Lambda$--enhanced gray molasses is performed for 10~ms on the $\mathrm{D}_1$ transition of \K{}, which reaches a final temperature of 20~\uK{}. After reloading the atoms into the quadrupole field, adiabatic compression raises their temperature to 100~\uK{}, and $1\times 10^9$ \Rb{} and $7\times 10^7$ \K{} are captured in the $|F,m_F\rangle = |2,2\rangle$ and $|9/2,9/2\rangle$ states, respectively.

Atoms are spatially transported $\sim$1~m to a high-vacuum science cell ($P \sim 10^{-11}$~torr) by physically translating the anti-Helmholtz coils producing the quadrupole field. The quadrupole trap is plugged with a blue-detuned beam (20~$\upmu$m waist, 760~nm), and magnetic evaporation is performed with a chirped driving of the $|2,2\rangle\to|1,1\rangle$ transition of \Rb{} using a microwave horn at 6.8~GHz. Potassium is sympathetically cooled, and at the end of magnetic evaporation, $6\times10^6$ of each species remain at 4~\uK{}. 

Once magnetic evaporation is complete, a crossed optical dipole trap (xODT) is turned on to capture the cold atoms, and the quadrupole field and plug beam are ramped off while a bias field of 30~G is turned on. The xODT is formed by two elliptical beams at 1064~nm with waists of $45 \times 170$~$\upmu$m crossing at $45^\circ$. Optical evaporation is performed by exponentially ramping the beams to variable cuts ($\sim$1/10 of their initial value) and then recompressing the trap such that the \K{} trap frequencies are $(\omega_x,\omega_y,\omega_z) = 2\uppi\times(45,250,80)$~Hz. Depending on the final trap depth, the atom number and temperature can be varied significantly. Typically, $10^6$ \Rb{} and $1.4\times 10^6$ \K{} atoms at 300~nK, or $7\times10^4$ \Rb{} and $5\times 10^5$ \K{} atoms at 50~nK can be produced.

During optical evaporation the atoms are transferred to the Feshbach states using adiabatic rapid passage (ARP). Rubidium is transferred from $|2,2\rangle \to |1,1\rangle$ and \K{} from $|9/2,9/2\rangle\to |9/2,-9/2\rangle$; each is transferred with about 98\% efficiency and untransferred atoms are blasted out of the trap with resonant light. Once optical evaporation and state transfer are complete, the bias field is ramped to 550~G in preparation for molecule production. The progression of atom numbers and temperature throughout the experimental cycle is shown in Fig.~\ref{fig:figureS1}.\\

\noindent\textbf{Production and imaging of ground state KRb.} In order to prevent gravitational sag from inducing oscillations in the ground state molecules, a vertical lattice formed by two counter-propagating beams (170~$\upmu$m waist, 1064~nm) is adibatically ramped on. It is found that a lattice depth of 30~$E_\mathrm{r}^\mathrm{KRb}$ is sufficient to suppress the effects of gravitational sag. Once molecules are produced, the lattice is ramped off in 5~ms. 

In the corrugated trap, weakly bound Feshbach molecules are produced by sweeping the magnetic field through the Fano-Feshbach resonance at 546.6~G. The field is swept from 556~G to 545.6~G in 3~ms. Conversion from unbound atoms to Feshbach molecules varies significantly depending on the initial temperature of the atoms, and it can be as high as 50\% at low temperature and as low as 15\% at high temperature. 

Molecules are transferred to the rovibronic ground state using stimulated Raman adiabatic passage (STIRAP). The two STIRAP lasers, which operate at 968~nm and 689~nm, are each locked to a common high-finesse optical cavity using the Pound-Drever-Hall method. The STIRAP sequence is 4~$\upmu$s long, and has a transfer efficiency of $\sim$90\%; the reported molecule numbers are corrected for this efficiency.

Immediately after molecule production, unpaired atoms are removed from the trap to mitigate molecule loss. To remove \K{}, a 30~$\upmu$s pulse of resonant light is applied to blast \K{} out of the trap while leaving the molecules unaffected. To remove leftover \Rb{}, which is in the $|1,1\rangle$ state, four ARP + blast sequences are applied to ensure the total removal of Rb. The molecule number and temperature are found to be unaffected by the atom removal. 

\begin{figure}[ht]
\centering
\includegraphics[width=2.3in]{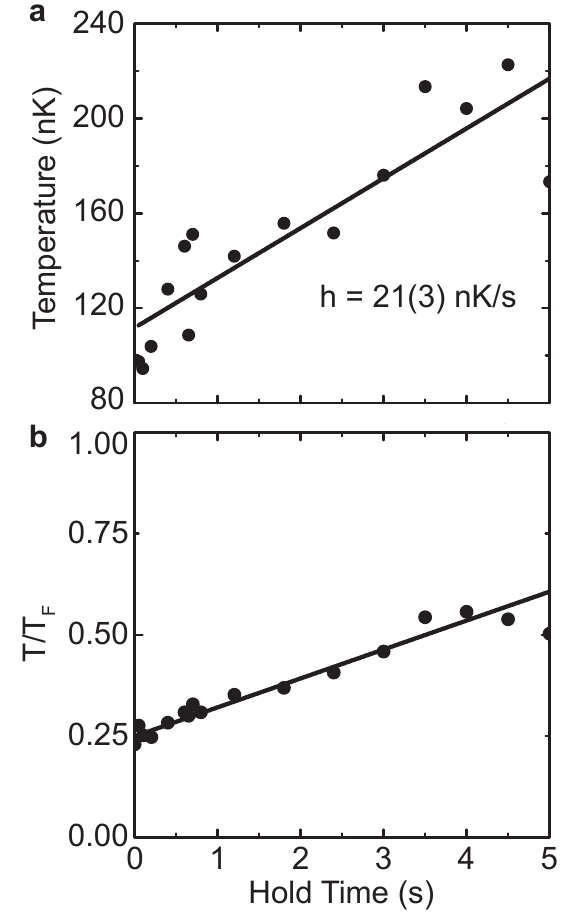}
\caption{\label{fig:figureS2} \textbf{a. Heating Rate.} The typical temperature of a molecular gas as a function of time. \textbf{b. Degeneracy Loss.} Despite the heating, \ttf{} remains small so long as $ht \ll T_0$. 
}
\end{figure}

To image molecules, the xODT is diabatically turned off and the STIRAP sequence is reversed to produced Feshbach molecules once again. The magnetic field is then jumped across the resonance to dissociate the weakly bound molecules into free atoms and the \K{} atoms are imaged on the $|9/2,-9/2\rangle\to |11/2,-11/2\rangle^\prime$ cycling transition after a variable amount of time of flight. We may also image Feshbach molecules directly without dissociation, by using resonant light to separate the molecules into free atoms. In this case, we must account for the reduced absorption cross section compared to free atoms, which we find to be about 0.7 times smaller.\\

\noindent\textbf{Image Fitting.} The molecular cloud is imaged from the side at an angle of $\theta = 22.5^\circ$ with respect to the principal axes of the trap. The imaging axes are
\begin{subequations}
\begin{align}
\hat{x}_1 &= \frac{\hat{x}\sin\theta + \hat{z}\cos\theta}{\sqrt{2}}\\
\hat{x}_2 &= \hat{y}.
\end{align}
\end{subequations}

The 2D column density distributions are fit to either a classical distribution for \ttf{}$\sim$1 or a Fermi-Dirac distribution for \ttf{}$\ll1$. 

For the classical distribution, we fit to 
\begin{equation}
n_\mathrm{Cl}(\hat{x}_1,\hat{x}_2) = n_0\mathrm{e}^{-\sum_{i=1}^2 \frac{\hat{x}_i^2}{2\sigma_i^2}} + c,
\end{equation}
with $n_0$, $\{\sigma_i\}$, and $c$ as fitting parameters.  Respectively, these are the peak density, the size of the cloud along each imaging axis, and the imaging offset. The cloud sizes are related to the classical temperature of the gas via
\begin{equation}
\sigma_i = \frac{\sqrt{1 + (\omega_i \tau)^2}}{\omega_i}\sqrt{\frac{k_\mathrm{B}T}{m}},
\end{equation}
where $k_\mathrm{B}$ is Boltzmann's constant, $m = 127$~amu is the molecular mass, $\omega_i$ is the trap frequency along the $i^\mathrm{th}$ direction, and $\tau$ is the time of flight.

For the Fermi-Dirac distribution, we make the Thomas-Fermi approximation and fit to 
\begin{equation}
n_\mathrm{FD}(\hat{x}_1,\hat{x}_2) = -n_0\mathrm{Li}_2\left(-z \mathrm{e}^{-\sum_{i=1}^2 \frac{\hat{x}_i^2}{2\sigma_i^2}}\right) + c.
\end{equation}  
Here, $z$ is the fugacity, which is a fitting parameter in addition to those described above. $\mathrm{Li}_s(z)$ is the polylogarithm function defined according to 
\begin{equation}
-\mathrm{Li}_s(-z) = \frac{1}{\Gamma(s)}\int_0^\infty \mathrm{d}x \frac{x^{s-1}}{z^{-1}\mathrm{e}^x + 1},
\end{equation}
where $\Gamma(s)$ is Euler's gamma function. The quantity \ttf{} is determined solely by the fugacity according to 
\begin{equation}
\left(\frac{T}{T_\mathrm{F}}\right)^3 = -\frac{1}{6\mathrm{Li}_3(-z)}.
\end{equation}

\noindent\textbf{Density loss fitting.} In a classical, harmonically trapped gas, the \textit{in situ} average density is given by 
\begin{equation}
n(T) =\frac{N}{V} = \frac{N}{8\uppi^{3/2}}\bar{\omega}^3\left(\frac{k_\mathrm{B}T}{m}\right)^{-3/2},
\end{equation}
where $\bar{\omega} = (\omega_1\omega_2\omega_3)^{1/3}$ is the geometric mean trap frequency. Differentiation of the above equation with respect to time yields Eq.~\ref{eq:densityrate}, with $\dot{N} = -\beta V n^2$ being the rate at which particles are lost. 

Since, according to the Bethe-Wigner threshold law, the two-body rate constant is proportional to temperature, it is convenient to write $\beta = bT$, with $b$ independent of temperature. Then, the rate equation reads
\begin{equation}\label{eq:densityrate_b}
\frac{\mathrm{d}n}{\mathrm{d}t} = -bTn^2 - \frac{3}{2}\frac{n}{T}\frac{\mathrm{d}T}{\mathrm{d}t}.
\end{equation}

While, in principle, anti-evaporation is the main source of heating, we observe a linear increase in temperature over the course of the hold time of the molecules; an example of which is shown in Fig.~\ref{fig:figureS2}a. We therefore, do not fit $T$, but measure it to be $T = T_0 + ht$, where $h$ is the heating rate and $t$ is the hold time. With this, Eq.~\ref{eq:densityrate_b} becomes
\begin{equation}\label{eq:densityrate_c}
\frac{\mathrm{d}n}{\mathrm{d}t} = -b(T_0 + ht)n^2 - \frac{3}{2}n\frac{h}{T_0 + ht}.
\end{equation}

The closed form solution for Eq.~\ref{eq:densityrate_c} is 

\begin{widetext}
\begin{equation}\label{eq:densitysolution}
n(t) = \frac{n_0 h T_0^{3/2}}{(ht+T_0)\left(2n_0 T_0^2\left(\sqrt{T_0}-\sqrt{ht+T_0}\right)b + h\left(\sqrt{ht + T_0} + 2n_0tT_0^{3/2}b\right)\right)},
\end{equation}
\end{widetext}

and density loss curves are fit to this equation. Example density loss curves and their corresponding fits are shown in Fig.~\ref{fig:figure3}b. 

In our analysis, $b$ and $n_0$ are fitting parameters, while $T_0$ and $h$ are measured. $T_0$ is measured at time $t=0$ by averaging 3--5 images and fitting to the Fermi-Dirac distribution, while $h$ is measured by considering the increase in $\sigma_i^2\propto T$ as a function of time. Despite the heating, molecular degeneracy is preserved over the course of the molecules' lifetime, as shown in Fig.~\ref{fig:figureS2}b. 

In Figs~\ref{fig:figure4}a and~\ref{fig:figure4}b, the quantities plotted are $\beta = bT_0$ and $b = \beta/T$, respectively.  

\subsection*{Density fluctuations in an ideal Fermi gas}

Within the Thomas-Fermi approximation, the density distribution of the ideal Fermi gas is given by
\begin{equation}
n(\bm{r}) = -\frac{1}{\Lambda^3}\mathrm{Li}_{3/2}\left(-z\mathrm{e}^{-\beta V(\bm{r})}\right).
\end{equation}
In the grand canonical ensemble, fluctuations in density are described in the thermodynamic limit by 
\begin{align}
\delta n(\bm{r})^2 &= \langle n^2(\bm{r})\rangle - \langle n(\bm{r})\rangle^2\nonumber\\
 &= \frac{1}{\beta} \frac{\partial n(\bm{r})}{\partial \mu}\nonumber\\
  &= -\frac{1}{\Lambda^3}\mathrm{Li}_{1/2}\left(-z\mathrm{e}^{-\beta V(\bm{r})}\right). 
\end{align}
Local relative number fluctuations are therefore described by
\begin{align}
\frac{\delta n(\bm{r})^2}{n(\bm{r})} = \frac{\mathrm{Li}_{1/2}\left(-z\mathrm{e}^{-\beta V(\bm{r})}\right)}{\mathrm{Li}_{3/2}\left(-z\mathrm{e}^{-\beta V(\bm{r})}\right)}.
\end{align}

However, in our experiment, we do not consider reactions locally, but rather globally. Therefore, it is necessary to average over all fluctuations in the trap. The normalized spatial probability distribution is given by 
\begin{align}
f(\bm{r}) &= \frac{n(\bm{r})}{N}\nonumber\\
&= \Lambda^3\left(\frac{m\bar{\omega}}{h}\right)^3\frac{1}{\mathrm{Li}_3(-z)}\mathrm{Li}_{3/2}\left(-z\mathrm{e}^{-\beta V(\bm{r})}\right),
\end{align}
so that the size of fluctuations averaged over the trap is 

\begin{equation}
\frac{\langle\delta n(\bm{r})^2\rangle}{\langle n(\bm{r})\rangle} = \frac{\int\mathrm{d}\bm{r}f(\bm{r})\delta n(\bm{r})^2}{\int\mathrm{d}\bm{r}f(\bm{r})n(\bm{r})}
\end{equation}

\begin{figure}[ht]
\centering
\includegraphics[width=2.3in]{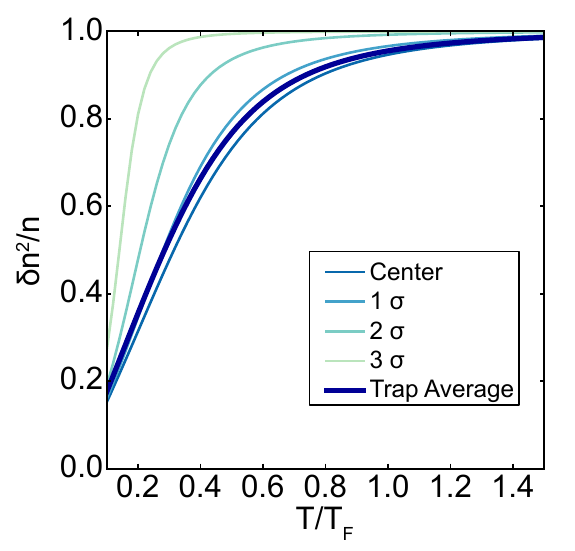}
\caption{\label{fig:fluc_supp} \textbf{Density Fluctuations in an ideal Fermi gas.} Relative density fluctuations in a harmonically trapped Fermi gas at various positions. The bold line corresponds to the average over the trap.  
}
\end{figure}

The average suppression of fluctuations is compared to the suppression of fluctuations in the center of the trap in Fig.~\ref{fig:fluc_supp}. It is also clear from Fig.~\ref{fig:fluc_supp} that fluctuations are only suppressed at the edge of the trap ($3\sigma$) for \ttf{}~$<0.2$, which is consistent with our expectation that chemical reactions are only strongly suppressed at the center. 

\end{document}